\documentclass[12pt]{article}
\usepackage{graphicx}
\usepackage{epsfig,cite}
\usepackage{amsmath,amssymb,graphicx,array,dcolumn,subfigure,rotating,color}
\usepackage{ae}
\usepackage{aecompl}
\usepackage{psfrag}

\newcommand{\Av}[1]{{\bf #1}}

\newcommand {\bnabla} {\mbox{\boldmath$\nabla$}}

\def\ln{{\operatorname{ln}}}

\def\Tr{{\operatorname{Tr}}}

\def\ch{{\operatorname{ch}}}
\def\tg{{\operatorname{tg}}}

\def\arctg{{\operatorname{arctg}}}

\begin{document}

\title{Electrostatic image effects for counter-ions between charged planar walls}
\author{M. Kandu\v c $^{1}$ and R. Podgornik $^{1,2}$\\
$^1$ Department of Theoretical Physics, \\
J. Stefan Institute, SI-1000 Ljubljana, Slovenia\\
$^2$ Department of Physics, Faculty of Mathematics and Physics, \\
University of Ljubljana, SI-1000 Ljubljana, Slovenia}
\maketitle

\begin{abstract} We study the effect of dielectric inhomogeneities on the interaction between two planparallel charged surfaces with oppositely charged mobile charges in between. The dielectric constant between the surfaces is assumed to be different from the dielectric constant of the two semiinfinite regions bounded by the surfaces, giving rise to electrostatic image interactions. We show that on the weak coupling level the image charge effects are generally small, making their mark only in the second order fluctuation term. However, in the strong coupling limit, the image effects are large and fundamental. They modify the interactions between the two surfaces in an essential way. Our calculations are particularly useful in the regime of parameters where computer simulations would be difficult and extremely time consuming due to the complicated nature of the long range image potentials.

\end{abstract}

\section{Introduction} 
Processes involving electrostatic interactions often play a dominant role in biological and soft-matter systems \cite{Andelman}. In aqueous environments charges on macromolecular surfaces such as membranes, self-assembled micelles, globular proteins and fibrous polysaccharides tend to dissociate and affect a wealth of functional, structural and dynamical properties \cite{Andelman}. Due to the nature of the most common solvent in these systems - water - the interfaces between  solvent and macromolecules are not only characterised by charges but also by a relatively large differences in respective dielectric properties.  In this contribution we will  study the effect of these dielectric inhomogeneities on the interactions between charged macromolecular surfaces mediated by the dissolved mobile counterions.  

Understanding the behaviour of charged systems starts with  the proper analysis of the counter-ion distribution around charged object. The traditional approach to Coulomb fluids has been the mean-field Poisson-Boltzmann (PB) formalism applicable  at weak surface charges, low counter-ion valency and high temperature \cite{Naji}. The limitations of and corrections to this approach become practically important in highly-charged systems where  counterion-mediated interactions between charged bodies start to deviate substantially from the mean-field accepted wisdom \cite{hoda}. One of the most important recent advances in this field has been the systematization of these non-PB effects based on the notions of {\sl weak} and {\sl strong} coupling approximations as championed by Netz and coworkers \cite{Naji}. These two approximations allow for an explicit and exact treatment of charged systems at two disjoint limiting conditions whereas the parameter space in between can only be analysed approximately and is mostly accessible only {\sl via } computer simulations. They also add a transparent and useful systematization to different approaches tending to upgrade or supplement the traditional PB way \cite{shklovskii}.

Both the weak and the strong coupling approximations are based on a functional integral or field-theoretic representation \cite{podgornik} of the grand canonical partition function of a system composed of fixed surface charges with intervening mobile counterions, and depend on the value of a single dimensionless coupling parameter $\Xi$. If the charge of the counterions with valency $q$ is $q e_0$, the Bjerrum length $l_B=e_0^2/4\pi\varepsilon\varepsilon_0kT$ is the length which measures the distance at which two unit charges interact with thermal energy $kT$ (in water at room temperature, the value is $l_B\approx 0.7$ nm) and the surface charge density of the bounding interfaces is $\sigma$, then the coupling parameter is given by ${\Xi}= 2\pi q^3 l_B^2 \sigma/e_0 \sim q^3 \sigma T^{-2}$. For later use let us also introduce the standard Gouy-Chapman length defined as $\mu= e_0 / 2\pi q l_B \sigma$ that measures the extent of the counter-ion layer next to a charged surface \cite{Andelman}. The meaning of the coupling parameter can now be easily understood by considering  the mean distance between counter-ions in the layer next to a charged interface. The volume available per ion is $4\pi a^3/3= qe_0\mu/\sigma$, so that the mean distance between counterions is of the order of $a= \mu (3{\Xi}/2)^{1/3}$. Since the height of the counter-ion layer is of order of the Gouy-Chapman length, it follows that in the weak-coupling case, defined by ${\Xi}\ll 1$, the height of the layer is much larger than the separation between two neighbouring counter-ions and thus the counterion layer behaves basically as a 3D gas. Each counter-ion in this case interacts with many others and the collective mean-field approach {\sl \' a la} Poisson-Boltzmann is completely justified. On the other hand in the case of the strong coupling limit, defined by ${\Xi}\gg 1$, the mean distance between counterions $a$ is much larger than the layer height, meaning that the counter-ion layer behaves as a 2D gas \cite{Netz}. In this case the mean-field approach breaks down, each counter-ion moving almost independently from the other ones along the direction perpendicular to the wall and the collective effects defining a mean-field being non-existent. The two limits are thus characterised by a low/high valency of the counterions and/or a small/large value of the surface charge density. The range of validity of both limits has been explored thoroughly in \cite{Netz}.

Formally it can be shown that the weak coupling limit can be straightforwardly identified with the saddle-point approximation of the field theoretic representation of the grand canonical partition function \cite{podgor}, and is reduced to the PB theory to the lowest order, the next one being due to harmonic fluctuations around the saddle-point that lead to a generalized zero-frequency Lifshitz term \cite{Parsegian}. The strong coupling approximation has no PB-like correlates \cite{strong}  since it is formally equivalent to a single particle description, corresponding to two lowest  order terms in the virial expansion of the grand canonical partition function. The consequences and the forms of these two limits of the Coulomb fluid description have been explored widely and in detail (for reviews, see \cite{Naji,hoda}).

In what follows we propose to add another aspect of the weak-strong coupling dichotomy but this time modified by the presence of dielectric inhomogeneities at the charged boundaries. These are ubiquitous in soft- and bio-systems since water, being the most common solvent, has a vastly different static dielectric constant ($\sim ~80$) from all matter composing (bio) macromolecules ($\sim ~2-5$). Every macromolecule - water interface is thus a source of a huge dielectric jump. As the presence of dielectric discontinuities leads to (Kelvin) electrostatic image interactions, it is exactly those that we intend to include in the analysis of the counterion mediated interactions between charged macromolecular surfaces. In this work we will delimit ourselves solely to the planparallel geometry exemplifying the case of interacting planar lipid membranes. Other geometries pertinent to interactions between {\sl e.g.} cylindrical DNA molecules will follow in a separate publication. 

Below, we will start {\sl \' a la} Netz and proceed {\sl via} the weak and the strong coupling limits all the while  consistently incorporating dielectric discontinuities at all levels of analysis. In the weak coupling limit that we address in order to present a complete and consistent analysis, we will show that for this particular choice of geometry the mean-field term in the interaction pressure is not affected by image interactions; they only transpire in the second-order fluctuation term around the mean-field. Since this term in the weak coupling limit is by necessity small, so is the image effect. The opposite is true for the strong coupling limit. Being effectively a single particle description it is quite susceptible to dielectric inhomogeneities and we will show that their effect on the interaction pressure is substantial. It can even add qualitatively new features to the interaction pressure that are absent when the dielectric discontinuities are ignored. The analytical results that we derive are particularly {\sl \' a propos} in the strong coupling limit since in this case even simulations
are hampered down by the long computer times needed to evaluate infinite sums of dielectric images in the energy function.

\section{The model}

\begin{figure}[!ht]
\centerline{\psfig{figure=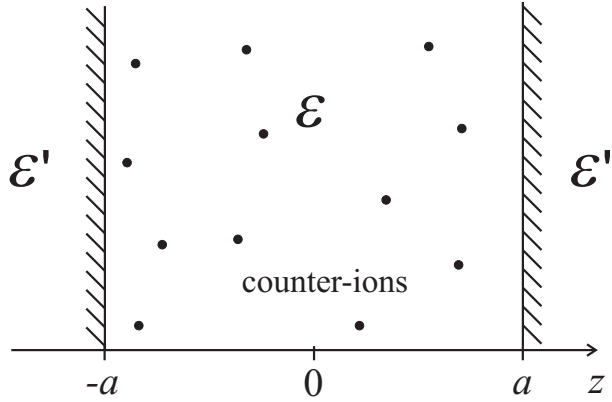,width=8cm}}
\caption{Counter-ions in a medium with an inhomogeneous dielectric constant confined between two charged
walls separated by $2a$. The space in between the surfaces has the dielectric constant $\varepsilon$, while the half-spaces outside the charged walls have a dielectric constant $\varepsilon'$. The fixed charge density of this system is given by $\sigma\delta(a-z)+\sigma\delta(a+z)$, where $\sigma$ is the surface charge density assumed to be equal at both bounding surfaces and being of the sign opposite to the sign of the counterions.}
\label{skica}
\end{figure}

Consider a system of $N$ counter-ions with valency $q$ confined between two oppositely charged walls with uniform charge distribution of surface charge density $\sigma$. Dielectric constant in the slab between the walls is $\varepsilon$ ({\sl e.g.} water) whereas outside it is assumed to be different and is denoted by $\varepsilon'$ ({\sl e.g.} hydrocarbons),  see Fig. \ref{skica}. For simplicity we set the origin of the coordinate system at the middle , so that the walls are positioned symmetrically at $z=-a$ and $z=a$.

Counter-ions interact {\sl via} Coulomb interaction potential $u(\Av r,\Av r')$ that is composed of the direct interaction $u_0(\Av r,\Av r')$ and the (electrostatic) image interaction $u_{\rm im}(\Av r,\Av r')$, being a consequence of the dielectric constant discontinuities at the boundaries of the system, thus
\begin{equation}
u(\Av r,\Av r')=u_0(\Av r,\Av r')+u_{\rm im}(\Av r,\Av r').
\label{imageterm}
\end{equation}
The 2D Fourier transform in lateral coordinates $x, y$ of the two interaction potentials is (see {\sl e.g.} \cite{schwinger})
\begin{eqnarray}
u_0(Q,z;z')&=&\frac{1}{2\varepsilon\varepsilon_0 Q}\,e^{-Q\vert z-z'\vert},\nonumber\\
u_{\rm im}(Q,z;z')&=&\frac{1}{\varepsilon\varepsilon_0 Q}\,
\frac{\ch\,Q(z+z')+\Delta\,e^{-2Qa}\,\ch\,Q(z-z')}{\Delta^{-1}e^{2Qa}-\Delta\, e^{-2Qa}},
\end{eqnarray}
where $Q$ is the magnitude of the 2D wave vector $\bf Q$. The partition function $Q_N$ of this Coulomb fluid composed of $N$ charged particles is given by
\begin{equation}
Q_N=\frac{1}{N!}\int d^3\Av r_1\ldots d^3\Av r_N\>
\exp\Bigl( - \beta U[\Av r_i,\Av r_j] \Bigr),\\
\end{equation}
with
\begin{eqnarray}
\beta U[\Av r_i,\Av r_j] &=& \frac{1}{2}\beta\sum_{i\ne j}e_0^2 q^2\,u(\Av r_i,\Av r_j) +
\beta\sum_i \int e_0q\,u(\Av r_i,\Av r)\rho_0(\Av r)d^3\Av r + \nonumber\\
& & + \frac{1}{2}\beta \int\!\!\!\int u(\Av r,\Av r')\rho_0(\Av r)\rho_0(\Av r) d^3\Av r d^3\Av r',
\end{eqnarray}
where $\rho_0(\Av r)=\sigma\delta(a-z)+\sigma\delta(a+z)$ stands for the fixed charge density of the two walls.
The three terms of $\beta U[\Av r_i,\Av r_j]$ correspond to direct electrostatic interaction between counterions, electrostatic interactions between counterions and fixed charges and between fixed charges on the walls themselves, respectively. 

We now proceed in the Netz \cite{Netz} fashion and perform the Hubbard-Stratonovitch transformation of the partition function 
\cite{podgor}, where the configurational integral over counterion positions is transformed into a functional integral over a fluctuating field $\phi$, which can be interpreted as the local fluctuating electrostatic potential. In this way the grand-canonical partition function is obtained straightforwardly in the form \cite{NajiNetz}
\begin{equation}
Z_G=\sum_{N=0}^\infty \lambda^N Q_N = C\int{\cal D}[\phi(\Av r)]\,e^{-\beta H[\phi(\Av r)]},
\label{funint}
\end{equation}
where $\lambda$ is the bare fugacity which is the exponential of the chemical potential. The prefactor $C$ above is the functional determinant of the inverse interaction potential $u^{-1}(\Av r,\Av r')$ while the "action" of the functional integral is given by
\begin{eqnarray}
\beta H[\phi]={\textstyle\frac{1}{2}}\beta\!\!\int\!\!\!\!\!\int\!\!\epsilon(\Av r) 
(\bnabla\phi(\Av r))^{2}d^3\Av r -i\beta\!\!\int\!\!\rho_0(\Av r)\phi(\Av r)d^3
\Av r-\lambda'\!\!\int\!\!e^{i\beta e_0 q\phi(\Av r)}d^3\Av r.
\label{action}
\end{eqnarray}
Here we have denoted the renormalized fugacity as $\lambda'=\lambda\,\exp(\frac{1}{2}e_0^2 q^2 u_0(\Av r,\Av r))$.  The functional integral Eq. \ref{funint} can be further rescaled \cite{NajiNetz} yielding 
$Z_G \longrightarrow C\int{\cal D}[\phi(\Av r)]\,e^{- \frac{H'[\phi(\Av r)]}{\Xi}}$, with dimensionless $H'[\phi(\Av r)]$, showing explicitly the dependence on the coupling parameter $\Xi$. The difference between the case with and without dielectric discontinuities is thus only in the spatial dependence of the dielectric constant indicated by $\epsilon(\Av r)$.  In this field-theoretical representation the grand canonical partition function is now ready to be evaluated explicitely and analytically  in the limiting cases of the weak and the strong coupling limits, as defined in the previous section \cite{Netz,Naji}.

\section{Weak coupling limit} 

In the limit of $\Xi \ll 1$ the functional integral Eq. \ref{funint} is dominated by the saddle-point value of 
the fluctuating potential, $\phi_0(\Av r)$, set by the extremal value of $H[\phi]$ \cite{podgor,Netz, NajiNetz}.
This is defined as a solution of 
\begin{equation}
\left(\frac{\delta H}{\delta \phi(\Av r)}\right) _{\phi_0} = 0.
\label{saddle}
\end{equation}
The lowest order correction to the saddle-point is obtained by expanding $H[\phi]$ around the extremum $\phi_0(\Av r)$ to the second order, the first order being zero by definition. To the lowest order in the deviation from the saddle-point fluctuating potential 
\begin{eqnarray}
H[\phi(\Av r)] = H[\phi_0(\Av r)] +  \frac{1}{2}\int\!\!\!\int \left(\frac{\delta^2 H}{\delta\phi(\Av r)\delta\phi(\Av r')}\right)_{\phi_0}
\delta \phi(\Av r)\delta \phi(\Av r')\,d^3\Av r\,d^3 \Av r'+{\cal O}(\delta\phi^3),\nonumber\\
~
\end{eqnarray}
with $\delta\phi(\Av r)=\phi(\Av r)-\phi_0(\Av r)$. The second order term obviously describes fluctuations around the mean-field.  By making a Wick rotation in the saddle-point potential, $\phi_0 \longrightarrow i\psi_0$, and writing Eq. \ref{saddle}
explicitly, we obtain 
\begin{equation}
\nabla^2\psi_0=-\frac{\lambda' e_0 q}{\varepsilon\varepsilon_0}\,e^{-\beta e_0 q\psi_0(\Av r)}.
\label{PB}
\end{equation}
This is obviously nothing but the mean-field PB equation, as first shown in \cite{podgornik}. The
saddle-point {\sl ansatz} thus corresponds exactly to the mean-field approximation of classical statistical mechanics of Coulomb fluids. In the case of surface distribution of external charges, as in the case of system Fig. \ref{skica}, the PB equation has to be supplemented by an appropriate boundary condition at $z = \pm a$ corresponding to the electroneutrality of the system. The Hessian of the Hamiltonian can now be evaluated explicitly and the corresponding grand canonical partition function for the mean-field plus the second order correction to the mean-field  can be  obtained straightforwardly as \cite{podgor,Netz}
\begin{equation}
Z_G = C\,\exp\Bigl(-\beta H[\phi_0] -{\textstyle\frac{1}{2}}\,\Tr\,\ln[u^{-1}(\Av r,\Av r')+\lambda'\beta e_0^2 q^2\, e^{-\beta e_0 q\psi_0(\Av r)}\delta^3(\Av r-\Av r')]\Bigr).
\label{ZGexpand}
\end{equation}
The second order fuctuational term can be shown to be equivalent to the first order loop expansion term \cite{Netz}. As already noted, the second order fluctuational correction to the mean-field presents a generalization of the zero-frequency Lifshitz interaction term \cite{podgornik}. In what follows we shall now evaluate Eq. \ref{ZGexpand} explicitly for the model system Fig. \ref{skica}.

\subsection{Mean-field}

The solution of the Poisson-Boltzmann equation Eq. \ref{PB} in the geometry of Fig. \ref{skica} is well known (see {\sl e.g.} Ref. \cite{Andelman}) being given by $\psi_0(z)=\frac{2}{\beta e_0q}\,\ln\,\cos\,\alpha z$ with $\alpha^2=\lambda'\beta e_0^2 q^2/2\varepsilon \varepsilon_0$, that can be determined from boundary conditions at the two bounding surfaces $z = \pm a$. They are of the form: $\alpha \,\tg\,\alpha a= 1/\mu$, where $\mu$ is the  Gouy-Chapman length. From the saddle-point solution one can obtain the corresponding grand canonical theormodynamic potential  $q=-kT\,\ln\,Z_G$. It implies a free energy defined by Legendre transform  $F=q+\tilde{\mu}N$, where $\tilde\mu$ represents the chemical potential $\tilde{\mu}=kT\,\ln\,\alpha^2$. Expressing all quantities in dimensionless units, so that $\tilde a=a/\mu$, $\tilde\alpha=\alpha/\mu$ and $\tilde p=p/(\sigma^2/2\varepsilon\varepsilon_0)$  the mean-field free energy is obtained as \cite{Andelman}
\begin{equation}
\tilde F_0/\tilde S=2\tilde\alpha^2 \tilde a+2\,\ln\left(\tilde \alpha^2 +1\right).
\label{free0}
\end{equation}
Here we have discarded all terms that do not depend on the intersurface separation $a$ since we are interested solely in the interactions between the two charged surfaces. In the above free energy we also omitted a term that would go as the logarithm of $C$ in the grand canonical partition function Eq. \ref{ZGexpand}. This term
actually leads to an additive contribution to the free energy  Eq. \ref{free0} that is however exactly cancelled by the zero-frequency Lifshitz - van der Waals term for two dielectric half-spaces that one has to add to the total free energy in a consistent treatment of the interactions (for details see \cite{podgornik,attard}).

The pressure corresponding to the free energy Eq. \ref{free0} and acting between the two surfaces  is obtained simply {\sl via} a derivative of the free energy with respect to intersurface volume, $p_{0}=-1/S\,( \partial F_{0}/ \partial (2a))$. In dimensionless form it can be written as \cite{Netz}
\begin{equation}
\tilde p_{0}=\tilde\alpha^2\approx
\left\{
\begin{array}{ll} 1/\tilde a-1/3&\textrm{ for }\tilde a\ll 1,\\
\pi^2/4\tilde a^2&\textrm{ for }\tilde a\gg 1.\\
\end{array}
\right.
\end{equation}
We see that on the mean-field level the pressure is always repulsive and does not depend on the dielectric
discontinuities in the system. This is a simple consequence of the fact that the mean-field solution depends {\bf only} on the longitudinal coordinate $z$, whereas image effects always depend on the induced charge density at the interface that necessarily has also transverze, {\sl i.e.} $(x,y)$, dependence. On a consistent mean-field level one thus can {\sl not} expect any image effects.

\subsection{Second order correction} 

We now turn to the second order expansion term, {\sl i.e.} the second term in Eq. \ref{ZGexpand}. We will deal with it only briefly since it was already derived before (see Refs. \cite{podgornik,attard}). From
the saddle-point solution we first of all get  
\begin{equation}
\lambda'\beta e_0^2 q^2\, e^{-\beta e_0 q\psi_0(\Av r)} = \varepsilon\varepsilon_0\,\frac{2\alpha^2}{\cos^2
\alpha z}.
\end{equation}
The tracelog in Eq. \ref{ZGexpand} can be written equivalently in the following form \cite{podgornik,attard}
\begin{eqnarray}
\Tr\,\ln[u^{-1}(\Av r,\Av r')+\lambda'\beta e_0^2 q^2\, e^{-\beta e_0 q\psi_0(\Av r)}\delta^3(\Av r-\Av r')]= 
\frac{S}{2\pi}\int_0^{\infty} Q\>\ln\frac{{\cal D}_1( Q)}{{\cal D}_0(Q)}\,dQ.
\label{trace}
\end{eqnarray}
The discrete sum of eigenvalues of the operator $u^{-1} + V$ has been replaced by an integral over the transverze wave-vector $\Av Q = (Q_{x}, Q_{y})$ with density  of modes $S/(2\pi)^2$, due to the translational homogeneity of the system in the transverse directions, and ${\cal D}_\lambda$ is the secular determinant of the fluctuating modes. The above formula is obtained by using the argument principle \cite{attard}.  The index $\lambda$ here refers to the eigenvalue equation  that can be derived in the form
\begin{equation}
\Bigl(\frac{
\partial^2}{
\partial z^2}-Q^2-\lambda\frac{2\alpha^2}{\cos^2\alpha z}\Bigr) f_{\lambda}(\Av Q,z)=0.
\label{eigF}
\end{equation}
The linearly independent solutions of this equation  for $\lambda=1$ inside the region $z \leq \vert{a}\vert$ are $y_{1}(z)=e^{Qz}\Bigl(1+\frac{\alpha}{Q} \,\tg\,\alpha z\Bigr)$ and  $y_{2}(z)=\frac{e^{-Qz}}{Q^2+\alpha^2}\Bigl(1-\frac{\alpha}{Q}\,\tg\,\alpha z\Bigr)$. For the external region $z > \vert{a}\vert$ the solutions are simply exponential functions $e^{\pm Qz}$. The secular determinant is then obtained in the following  form
\begin{equation}
D_1(Q)=g(Q)\left(1-\Delta^2(Q)\,e^{-4Qa}\right),
\end{equation}
with
\begin{eqnarray}
g(Q)&=&\frac{\left[(1+\Delta)(1+\alpha^2\mu^2)+ 2Q\mu+2Q^2\mu^2\right]^2}{Q^2+\alpha^2},\nonumber\\
\Delta(Q)&=&\frac{(1+\Delta)(1+\alpha^2\mu^2)+2\Delta(Q^2\mu^2-Q\mu)} {(1+\Delta)(1+\alpha^2\mu^2)+2(Q^2\mu^2+Q\mu)}.
\end{eqnarray}
The dielectric jump at the boundaries of the system $z = \pm a$ is characterised by $\Delta=(\varepsilon-\varepsilon')/ (\varepsilon+\varepsilon')$. We can now finally write down the second order correction term $F_2$ to the free energy in dimensionless units as
\begin{equation}
\tilde{F}_2/\tilde{S}={\textstyle\frac{1}{2}}{\Xi}~\Bigl[
\int_0^{\infty}\tilde{Q}\,\ln\,g(\tilde Q)\,d\tilde Q+
\int_0^{\infty}\tilde{Q}\,\ln\bigl(1-\Delta^2(\tilde Q)\, e^{-4\tilde Q\tilde a}\bigr)d\tilde Q\Bigr].
\label{finale-1}
\end{equation}
which is obviously proportional to the coupling parameter $\Xi$.
In order to extract the finite contributions to the free energy, we have to regularize the first integral in the above expression  that yields 
\begin{equation}
\int_0^{\infty}\tilde{Q}\,\ln\,g(\tilde Q)\,d\tilde Q=
\frac{1}{2}\tilde\alpha^2(\Delta+\ln\,2\tilde\alpha^2)-h(\tilde\alpha)\,\arctg\,h(\tilde\alpha)
-\frac{h^2(\tilde\alpha)-1}{4}\,\ln\,\frac{h^2(\tilde\alpha)+1}{2},
\end{equation}
where
\begin{equation}
h(\tilde\alpha)=\sqrt{1+2\Delta+2(1+\Delta)\tilde\alpha^2}.
\end{equation}
The second integral in Eq. \ref{finale-1} is convergent and can be straightforwardly evaluated numerically. 

Before actually proceeding numerically we first consider the limits of small and large spacings analytically.
The second-order free energy Eq. \ref{finale-1} allows for a simple asymptotic expression valid for small 
$\tilde a$ of the form
\begin{equation}
\tilde{F}_2/\tilde{S}\sim-{\Xi}\,\frac{\textrm{Li}_3(\Delta^2)}{32\,\tilde a^2}.
\end{equation}
Here, $\textrm{Li}_3$ is a polylogarithm function. This contribution comes from the second
term of Eq. \ref{finale-1} and is valid for $\Delta\ne 0$. The first term of Eq. \ref{finale-1} contributes only to the order $1/\tilde a$ and is therefore not inculded in the asymptotic expression. The corresponding pressure is obtained from the derivative of the free energy with respect to the intersurface volume, and if we combine it with the mean-field expression we get for the total pressure in the $\tilde a\ll 1$ limit
\begin{equation}
\tilde p = \tilde p_0 + \tilde p_2\sim \frac{1}{\tilde a}-{\frac13} -{\Xi}\,\frac{\textrm{Li}_3(\Delta^2)}{32\,\tilde a^3}.
\end{equation}
Here one needs to remember that the whole approach, meaning mean-field plus quadratic fluctuations around the mean-field,  only works if the fluctuation contribution is a small perturbation to the dominating mean-field contribution. Thus the above formula can be assumed correct only if the last term is smaller than the first two.
Note that only the fluctuation term depends on the dielectric discontinuity.

In the opposite limit of large $\tilde a$, the asymptotic form of the dimensionless second order free energy can be derived in the form
\begin{eqnarray}
\tilde{F}_2/\tilde{S}&\sim&-\frac{\Xi}{32\,\tilde a^2}\Bigl[2\pi^2\,
\Bigl(\ln\,\frac{2\tilde a^2}{\pi^2}+
1+2\frac{1+\Delta}{\sqrt{1+2\Delta}}\,\arctg\sqrt{1+2\Delta}-\nonumber\\
&&-(1+\Delta)\ln(1+\Delta)\Bigr)
+\zeta(3)\Bigr].
\end{eqnarray}
The largest contribution goes as $\ln\,\tilde a\,/\tilde a^2$ and does not involve image effects, {\sl i.e.} is independent of $\Delta$. Image contribution can only be discerned at the order  $1/\tilde a^2$ and its
magnitude increases with increasing $\Delta$. The asymptotic expression for the total pressure for $\tilde a\gg 1$ is then
\begin{eqnarray}
\tilde p = \tilde p_0 + \tilde p_2&\sim& \frac{\pi^2}{4\tilde a^2} -\frac{\Xi}{32\,\tilde a^3}\Bigl[2\pi^2\,
\Bigl(\ln\,\frac{2\tilde a^2}{\pi^2}+
2\frac{1+\Delta}{\sqrt{1+2\Delta}}\,\arctg\sqrt{1+2\Delta}-\nonumber\\
&&-(1+\Delta)\ln(1+\Delta)\Bigr)
+\zeta(3)\Bigr].
\end{eqnarray}
In this limit the fluctuation terms are consistently smaller than the mean-field term. Again only the fluctuation term depends on the dielectric discontinuity.

\subsection{Numerical results}
\begin{figure}[!ht]
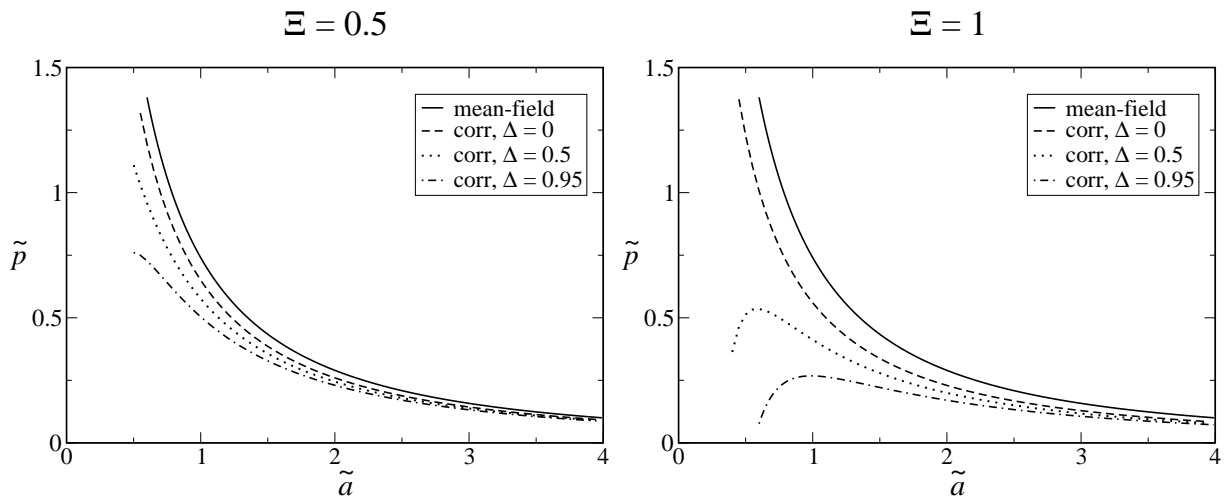

\centerline{\psfig{figure=press-wc05.eps,width=8cm}
\psfig{figure=press-wc1.eps,width=8cm}}
\caption{Rescaled pressure between the charged walls located at $z = \pm a$, as a function of the rescaled half-distance $\tilde a$. The mean-field approximation is compared with the total pressure involving the second-order correction for various dielectric jumps $\Delta$ and for coupling parameters ${\Xi}=0.5$ and $1$.}
\label{fig-pressWC}
\end{figure}
Numerical results for the total pressure acting between the walls, corresponding to the mean-field plus quadratic
fluctuations around the mean-field, and given in the dimensionless form by 
\begin{equation}
\tilde p(\tilde a) = \tilde p_{0}(\tilde a) + \tilde p_{2}(\tilde a),
\end{equation}
and are presented in Fig. \ref{fig-pressWC} as a function of dimensionless separation $\tilde a$. The second order correction $\tilde p_{2}(\tilde a)$ obviously gives an attractive contribution to total pressure. As
$\tilde p_{2}(\tilde a)$ becomes comparable to the leading order mean-field term $\tilde p_{0}(\tilde a)$ the loop-expansion is expected to break down. As a consequence, this method can not be used to predict a net attraction between two charged plates. Since only the fluctuation term, $\tilde p_{2}(\tilde a)$, depends on the dielectric discontinuity its effects are by necessity small in this limit. For large coupling parameters $\Xi$ a different, non-perturbative theory must be used to which we turn next.

\section{Strong coupling limit} 

In the strong coupling limit, where ${\Xi}\gg 1$, the saddle-point expansion used before does not work \cite{Naji}. Here, a straightforward perturbation expansion in terms of fugacity $\lambda$ can be used instead, corresponding formally to the virial expansion of the grand potential Eq. \ref{funint}. For our purposes only the first two terms of this expansion will be used, leading to
\begin{eqnarray}
Z_G&=&Z_G^{(0)}+\lambda'\>Z_G^{(1)}+{\cal O}(\lambda'^2).
\end{eqnarray}
Here the zeroth order term corresponds to the bare electrostatic interaction energy of charged walls without any
counter-ions, and the first order term corresponds to a one-particle contribution \cite{Netz}. By making an inverse
Hubbard-Stratonovitch transformation, the zeroth order term reduces to a very simple form
\begin{eqnarray}
Z_G^{(0)} &=&e^{-\frac{1}{2}\beta\int\!\!\!\int d^3\Av r d^3\Av r' u(\Av r,\Av r')
\rho_0(\Av r)\rho_0(\Av r')}.
\label{scZG0}
\end{eqnarray}
On the other hand, the first order term is a bit more complicated, but nevertheless leads straightforwardly to
\begin{equation}
Z_G^{(1)}=Z_G^{(0)}\int e^{-\beta e_0 q\int u(\Av r,\Av R)
\rho_0(\Av r)\,d^3 \Av r-\frac{1}{2}\beta e_0^2 q^2 u(\Av R,\Av R)} d^3\Av R.
\end{equation}
Note here that in both cases we are dealing with the complete electrostatic interaction potential, including the image term Eq. \ref{imageterm}.  For simplicity we now introduce $U$, so that the grand partition function can be written as $Z_G=Z_G^{(0)}(1+\lambda U)$, with
\begin{eqnarray}
U=\int \exp\left(-\frac{1}{2}\beta e_0^2q^2 u_{\rm im}(\Av R,\Av R)-\beta e_0 q\int u(\Av r,\Av R)\rho_0(\Av r)d^3 \Av r\right) d^3\Av R
\label{scU}
\end{eqnarray}
To calculate the pressure between charged walls, we now follow the same procedure as we did in the saddle-point approximation. The first step is to evaluate the grand potential $q$ from the grand partition function $Z_G$. Because of the relation $\lambda U\ll 1$ applicable in this limit, we can expand the corresponding logarithm and the grand potential can be simply written as
$q = q_0-kT\lambda U$, 
where we have also introduced the zeroth order grand potential $q_0=-kT\,\ln\,Z_G^{(0)}$. Free energy can again be
obtained by adding Gibbs free enthalpy $\tilde\mu N$ to the grand potential. Since free energy does not explicitly depend on chemical potential {\sl i.e.} fugacity, $\partial F/\partial\lambda=0$ leads to
a relation between the number of counter-ions and $\lambda U$ of the form $N=\lambda U$.
The free energy, up to an irrelevant constant, is finally obtained in the strong coupling limit as
\begin{equation}
F(a)=q_0(a)-NkT\,\ln\,U(a).
\label{scF}
\end{equation}

In order to evaluate the integrals $q_0$ and $U$, we can first integrate out the lateral coordinates, since our system is translationally homogeneous in the lateral directions. The zeroth order is then given by
\begin{eqnarray}
q_0 &=&{\textstyle\frac{1}{2}}\int\!\!\!\!\int dz\,dz'\,u(\Av Q=0,z;z')\rho_0(z)\rho_0(z').
\end{eqnarray}
Inserting the Green function $u(\Av Q,z,z')$ and the charge density of the walls $\rho_0(\Av
r)=\sigma\,\delta(z-a)+\sigma\,\delta(z+a)$ the above expression is transformed into
\begin{equation}
q_0=-\frac{Sa\sigma^2}{\varepsilon\varepsilon_0}\left(\frac{1+\Delta}{1-\Delta}
\right)^2.
\end{equation}
Here we have discarded all therms that do not depent on $a$ since again we are interested only in the interactions between the two charged walls. In the same fashion we evaluate also the two terms involved in expression $U$, Eq. \ref{scU}. The self-image term can be expanded in series of $\Delta$ and then integrated
\begin{eqnarray}
u_{\rm im}(\Av R,\Av R)&=&\int\frac{d^2\Av Q}{(2\pi)^2}\,u_{\rm im}(Q,z;z)\nonumber\\
&=&\frac{a}{4\pi\varepsilon\varepsilon_0}	\sum_{n\textrm{ odd}}\frac{n\Delta^n}{n^2a^2-z^2}-\frac{\ln\,(1-\Delta^2)} {8\pi\varepsilon\varepsilon_0 a}.
\end{eqnarray}
Here the sum is over all odd integers $n=1,3,5,\ldots$. The other term can be expressed in a much simpler form, namely
\begin{equation}
\int d^3\Av r\,u(\Av r,\Av R)\rho_0(\Av r)=-\frac{\sigma a}{\varepsilon\varepsilon_0}
\left(\frac{1+\Delta}{1-\Delta}\right)^2.
\end{equation}
The dimensionless free energy is finally obtained within the strong coupling approximation as
\begin{eqnarray}
\tilde F/\tilde S=2\tilde a\left(\frac{1+\Delta}{1-\Delta}\right)^2-
\frac{\Xi}{2\tilde a}\,\ln(1-\Delta^2)-2\,\ln \,\tilde a\int_{0}^{1} \exp\left[-\frac{{\Xi}}{2\tilde{a}} \sum_{n\textrm{ odd}}\frac{n\Delta^n}{n^2-t^2}\right]\,dt.\nonumber\\
~
\label{freeSC}
\end{eqnarray}
From this dimensionless free energy the dimensionless pressure is obtained analogously to the weak coupling case. We note here that in the case of no images, corresponding to a dielectrically homogeneous case and thus to $\Delta = 0$, the dimensionless pressure can be evaluated explicitly giving $\tilde p=1/\tilde a-1$, as shown in \cite{Naji,Netz}.

It is again interesting to  look at the asymptotic behaviour of pressure for small and large wall separations. However, one needs to exercise some caution here. Let us first assume that $\Delta>0$ and is finite. In the case of $\tilde a\ll 1$ the main contribution to the integral Eq. \ref{freeSC} comes from small values of the integration variable $t$. Therefore, we can expand the rational function of $t$ to the second order and obtain a Gaussian integral. The asymptotic expansion to the lowest order for $\tilde a\ll 1$ is then obtained as
\begin{equation}
\tilde F/\tilde S(\tilde a, \Delta>0) \sim
2\tilde a\left(\frac{1+\Delta}{1-\Delta}\right)^2-
\frac{\Xi}{\tilde a}\,\ln(1-\Delta) -3\,\ln\,\tilde a + {\cal O}(\tilde a^2),
\end{equation}
with the corresponding pressure
\begin{equation}
\tilde p(\tilde a, \Delta>0)\sim-\left(\frac{1+\Delta}{1-\Delta}\right)^2-
\frac{\Xi}{2\tilde a^2}\,\ln(1-\Delta)+\frac{3}{2\tilde a} + {\cal O}(\tilde a).
\label{pres-1}
\end{equation}
The interaction pressure obviously depends strongly on the value of the dielectric discontinuity. As $\Delta \longrightarrow 0$ one has to carefully pick the terms in Eq. \ref{freeSC} that remain finite, ending up with
\begin{equation}
\tilde F/\tilde S(\tilde a, \Delta \longrightarrow 0)= 2 (\tilde a - \,\ln\,\tilde a)
+ \bigl(8 \tilde a + \frac{\Xi}{\tilde a}\bigr)\Delta + {\cal O}(\Delta^2).
\end{equation}
with the pressure
\begin{equation}
\tilde p(\tilde a, \Delta \longrightarrow 0)= \frac{1}{\tilde a} - 1 +
\bigl(\frac{\Xi}{2\tilde a^2} - 4\bigr) \Delta + {\cal O}(\Delta^2).
\label{pres-2}
\end{equation}
Clearly the pressure for $\Delta = 0$ reduces to $\tilde p=1/\tilde a-1$, which is consistent with the result obtained with no images \cite{Naji,Netz}. The above two limits do not imply any discontinuity for pressure at $\Delta = 0$. The change from the asymptotics Eq. \ref{pres-1} to \ref{pres-2} is smooth and continuous in $\Delta$.

In the opposite asymptotic limit of large $\tilde a$, the exponential function in 
Eq. \ref{freeSC} is approximately $1$ and the free energy can be written straightforwardly as
\begin{equation}
\tilde F/\tilde S\sim 2\tilde a\left(\frac{1+\Delta}{1-\Delta}\right)^2-
\frac{\Xi}{2\tilde a}\,\ln(1-\Delta^2)-
2\,\ln\,\tilde a + {\cal O}(\tilde a^{-2}).
\end{equation}
The corresponding pressure is then
\begin{equation}
\tilde p\sim-\left(\frac{1+\Delta}{1-\Delta}\right)^2-\frac{\Xi}{4\tilde a^2}\,\ln(1-\Delta^2)+\frac{1}{\tilde a} + {\cal O}(\tilde a^{-3}).
\label{equnew-1}
\end{equation}
Again the dependence on the dielectric discontinuity is quite strong. The expression involving no dielectric discontinuities is here obtained simply by setting $\Delta = 0$ directly in Eq. \ref{equnew-1}.

Apart from the interaction pressure we also evaluate the counterion density in order to quantify the image effects on the distribution of mobile charges between the two surfaces. The counter-ion density can be evaluated following \cite{Netz} by using the relation $N=\lambda U$ and inserting the expression for $U$. The integrand can then be recognised as the counter-ion density of the form
\begin{equation}
\tilde{\rho}(\tilde z)=A(\tilde a)\,\exp\left[-\frac{\Xi \tilde a}{2}\sum_{n\textrm{ odd}}\frac{n\Delta^n}{n^2\tilde{a}^2-\tilde{z}^2}\right]
\label{rho}
\end{equation}
The normalization factor $A(\tilde a)$ can be obtained by the volume integration of density and the electroneutrality {\sl ansatz}.

At the end we note here, that we have to be aware of the range of validity of the strong coupling limit (${\Xi}\gg 1$) which is formally given by the Rouzina-Bloomfield criterion \cite{Rouzina} ${\Xi}>\tilde a^2$ as formulated by Netz \cite{Netz}. At a given coupling parameter the strong coupling limit thus applies only to sufficiently small
intersurface separation. Conversely at a given separation the strong coupling limit applies only to sufficiently high values of the coupling parameter.

\subsection{Numerical results} 

\begin{figure}[!ht]
\centerline{\psfig{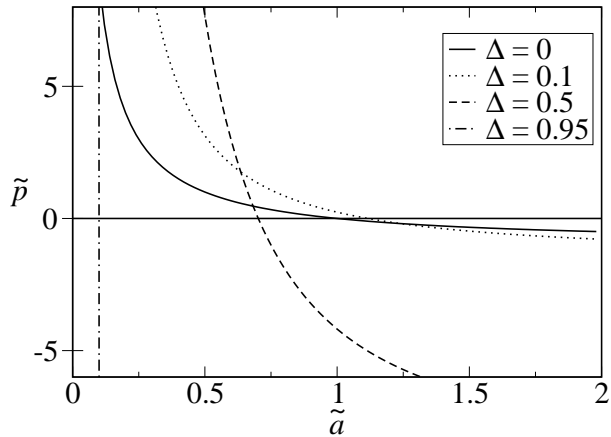}}
\caption{Rescaled interaction pressure between the charged walls as a function of the rescaled half-distance $\tilde a$ for coupling parameter ${\Xi}=10$ and various values of the dielectric jump defined by $\Delta$.}
\label{fig-pressSC}
\end{figure}

We now present numerical results for the dimensionless interaction pressure in the strong coupling limit.  It depends only on $\tilde a$ and $\Xi$.  In the general case of  nonvanishing $\Delta$ the pressure has to be evaluated numerically and is shown in Fig. \ref{fig-pressSC}.
It is obvious from figure Fig. \ref{fig-pressSC} that the rescaled pressure depends strongly on the value of the dielectric discontinuity $\Delta$. For larger $\Delta$ the attraction at larger separations as well as repusion at smaller separations are larger. The attractive pressure obviously diverges  as $\Delta$ approaches its limiting value of 1, valid for ideally polarizable surfaces.

\begin{figure}[!ht]
\centerline{\psfig{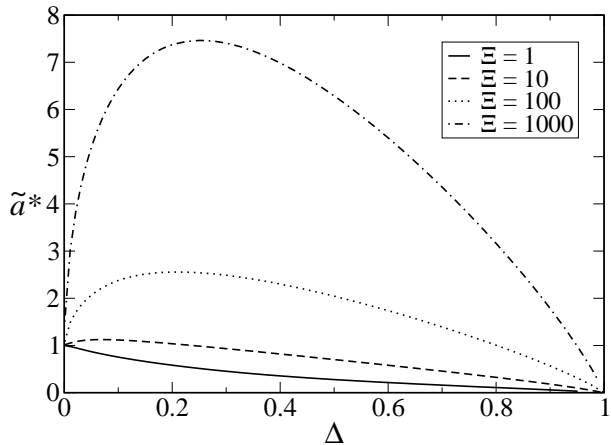}}
\caption{Rescaled half-separation between the walls of bound state as function of dielectric jump $\Delta$ for various values of coupling parameter $\Xi$.}
\label{fig-bound}
\end{figure}

We note that just as in the case of no dielectric inhomogeneities \cite{NajiNetz} there exists an intermediate value of the intersurface separation $\tilde a^*$ at which the pressure equals zero. This separation defines the {\sl bound state} of the system and has been studied extensively \cite{NajiNetz}. In the case of dielectric inhomogeneities the position of the bound state depends drastically on the value of $\Delta$, see Fig. \ref{fig-bound}. For $\Delta=0$ the bound state is at $\tilde a^*=1$ \cite{NajiNetz}. This value increases with increasing
$\Delta$ and reaches a maximum, then decreasing down to zero as $\Delta$ approaches 1.
This non-monotonic dependence of the bound-state on the dielectric discontinuity is quite interesting. In the case of an ideally polarizable interface in the strong coupling limit we would thus have complete collapse with $\tilde a^*=0$. Here the correlation effect obviously overwhelms the entropy of the conterions, collapsing the system completely. We just note here that for $\Delta<0$ the strong coupling pressure is purely attractive and the system would collapse completely for any $\Delta<0$.

While the dielectrically homogeneous system has been studied extensively and the analytic calculations can be
exhaustively compared with existant computer simulations this is not the case for dielectrically inhomogeneous
systems. The presence of image effects and the underlying image potential restricts the computer time considerably and the corresponding simulation studies are scarce.  The only available computer simulation data are the Monte Carlo simulations by Naji et al. \cite{Naji}.
\begin{figure}[!ht]
\centerline{\psfig{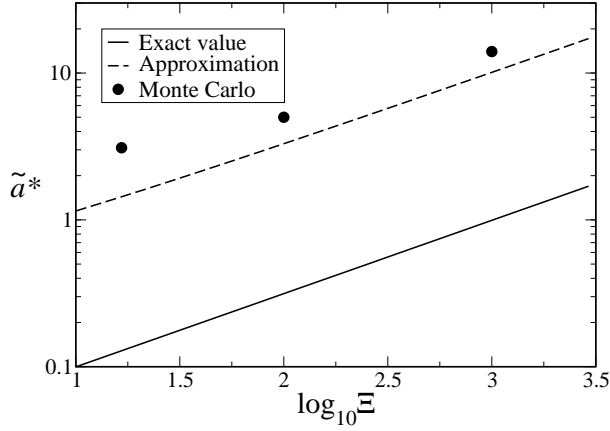}}
\caption{Bound state in terms of separation $\tilde a^*$ as a function of $\Xi$ at dielectric jump $\Delta=0.95$ at
both walls. Dashed line represents the linear approximation and circles correspond to Monte-Carlo simulations for first-order images \cite{Naji}.}
\label{boundXi}
\end{figure}
Unfortunately even these simulations can not be compared directly with our results. The main deficiency of these simulations is the approximate description of the image effects, since only the first-order
images are taken into account. Focusing on the bound-state wall separations for $\Delta=0.95$ (corresponding to water-hydrocarbon interface), the simulation results are an order of magnitude larger than our numerical results. The major part of discrepancy is surely caused by the first order approximation in the simulations. Namely, if we expand our expression for the free energy Eq. \ref{freeSC} to the first order in $\Delta$, thus making a similar approximation as in the simulations, our results are only about 30 \% off, as shown by dashed line in Fig. \ref{boundXi}. From this we can conclude that image contribution to the interaction is of crucial importance.
\begin{figure}[!ht]
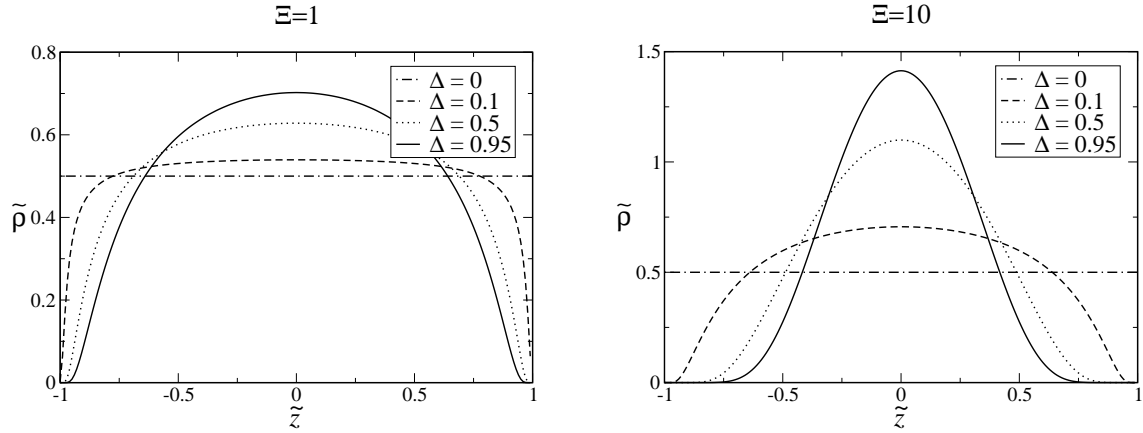

\centerline{\psfig{figure=dens-sc1.eps,width=7cm}\hspace{5ex}
\psfig{figure=dens-sc10.eps,width=7cm}}
\caption{Density profile between two charged walls separated by distance $2\tilde a=2$ for various dielectric jumps $\Delta$ at coupling parameters ${\Xi}=1$ and ${\Xi}=10$.   Here the volume integral is normalized to 1.}
\label{fig-density}
\end{figure}

At the end let us also analyze results for counter-ion density as shown in Fig. \ref{fig-density}. We note that in general the image effects change the distribution of counterions. For repulsive image interactions the counterions tend to be depleted in the vicinity of the charged surfaces. We note that this depletion appears to be weak for small dielectric jumps $\Delta$ and becomes significant for larger $\Delta$. This of course
makes sense, since higher $\Delta$ means larger charge of images on the other side of the wall that causes higher electrostatic repulsion, so counter-ions tend to concentrate toward the center. This effect also increases with increasing coupling constant $\Xi$. Obviously, the equation Eq. \ref{rho} is reduced to a constant when $\Delta$ vanishes, since in that case there is no image repulsion.

\section{Conclusions} 
In this paper we have analysed the effect of dielectric discontinuities on the interaction pressure between two charged surfaces with intervening counterions in planparallel geometry. In this analysis we have followed the approach of Netz and coworkers by explicitly considering the limits of weak and strong coupling.  This limits refer to the value of the electrostatic coupling parameter $\Xi$, which is assumed small in the first limit and large in the second, respectively.

Consider first the weak coupling limit with additive contributions from the mean-field and harmonic fluctuations around the mean-field.  We  note that the mean-field term in the interaction pressure is not affected by image interactions. This is easy to understand: in planparallel geometry the mean field depends only on the transverse coordinate, while image interactions by necessity imply also dependencies in the transverse directions. Thus by construction the mean field solution in this geometry does not depend on the presence of electrostatic image interactions. One could add those in by hand, as was done recently by Onuki \cite{onuki}, but such an approach does not {\sl sensu stricto} correspond to a mean-field analysis. On the weak coupling level the image interactions thus  only transpire in the second-order fluctuation term around the mean-field and have the form of a generalized zero-frequency Lifshitz term. Since this term in the weak coupling limit is by necessity small, so is in general the image effect. 

The opposite is true for the strong coupling limit which is effectively a single particle description since it corresponds to the second order virial expansion. As such it is quite susceptible to dielectric inhomogeneities and their effect on the interaction pressure is substantial. The image effects boost the value of the interaction pressure in the repulsive - small separations - as well as attractive -larger separations - domain. For the infinitely polarizable case in fact the pressure can reach infinite values. Also the bound state, defined as the separation at which the interaction pressure is zero, depends drastically and non-monotonically on the value of the dielectric discontinuity. For large differences in the value of the dielectric constant the system shows a tightly bound state.

Our computation has an important practical aside. As we have shown, the first order image correction is not enough to describe accurately the dielectric discontinuity effects on the strong coupling level. Even simulations that take into account only the first order images are not particularly accurate and miss most of the image effect in the interaction pressure. Our calculation is thus particularly valuable in the region of parameter space where other methods obviously fail quite seriously.

The validity of our approach is set by the constraints of the weak-strong coupling formalism as well as its coupling to the image effects. This is particularly clear in the latter. The approach followed in this contribution only makes sense if effective lateral separation between counterions is larger than the separation between images in the longitudinal direction. If this is not the case the strong coupling limit as concieved of above could not be constructed.

\section{Acknowledgement}

RP would like to acknowledge the financial support by the Slovenian Research Agency under contract 
Nr. P1-0055 (Biophysics of Polymers, Membranes, Gels, Colloids and Cells). MK would like to acknowledge the financial support by the Slovenian Research Agency under the young researcher grant.

\end{document}